\documentclass[lettersize, journal]{IEEEtran}
\usepackage{amsmath,amsfonts}
\usepackage{algorithmic}
\usepackage{array}
\usepackage{breqn}
\usepackage{textcomp}
\usepackage{stfloats}
\usepackage{url}
\usepackage{verbatim}
\usepackage{graphicx}
\usepackage{cite}
\usepackage{amssymb}
\usepackage[algo2e]{algorithm2e}
\usepackage{subfigure}
\usepackage{wrapfig}
\usepackage{algorithm}
\usepackage{bbm}
\usepackage{bbold}
\usepackage{amsmath}
\usepackage{amsfonts}
\usepackage{amssymb}
\usepackage{amsthm}
\usepackage{mathrsfs}

\usepackage{color}
\usepackage{ulem}
\usepackage{multirow}
\usepackage[table,xcdraw]{xcolor}
\usepackage{setspace}

\begin{document}

\title{Crowdsense Roadside Parking Spaces with Dynamic Gap Reduction Algorithm}

\author{
\IEEEauthorblockN{Wenjun Zheng, Zhan Shi, Qianyu Ou,  and Ruizhi Liao}
\IEEEauthorblockA{}
\thanks{This work was supported in part by NSFC (61902332), Shenzhen Education Project (JYPJ22001), Shenzhen Stability Science Program (20231130170021001), Guangdong Provincial Key Laboratory of Mathematical Foundations for Artificial Intelligence (2023B1212010001), and SRIBD Doctoral Scholarship Program. Wenjun Zheng and Zhan Shi contributed equally to this work. (Corresponding author: Ruizhi Liao). 

Wenjun Zheng, Zhan Shi and  Qianyu Ou are with School of Data Science, The Chinese University of Hong Kong, Shenzhen, China  (e-mail: wenjunzheng@link.cuhk.edu.cn; zhanshi1@link.cuhk.edu.cn; qianyuou@link.cuhk.edu.cn).


Ruizhi Liao is with Guangdong Provincial Key Laboratory of Mathematical Foundations for Artificial Intelligence; Shenzhen Key Laboratory of IoT
Intelligent Systems and Wireless Network Technology; School of Humanities and Social Science, The Chinese University of Hong Kong, Shenzhen, China  (e-mail: rzliao@cuhk.edu.cn).}}


\IEEEpubid{0000--0000/00\$00.00~\copyright~2022 IEEE}

\maketitle

\begin{abstract}
In the context of smart city development, mobile sensing emerges as a cost-effective alternative to fixed sensing for on-street parking detection. However, its practicality is often challenged by the inherent accuracy limitations arising from detection intervals. 
This paper introduces a novel Dynamic Gap Reduction Algorithm (DGRA), which is a crowdsensing-based approach aimed at addressing this question through parking detection data collected by sensors on moving vehicles. The algorithm's efficacy is validated through real drive tests and simulations. We also present a Driver-Side and Traffic-Based Model (DSTBM), which incorporates drivers' parking decisions and traffic conditions to evaluate DGRA's performance. Results highlight DGRA's significant potential in reducing the mobile sensing accuracy gap, marking a step forward in efficient urban parking management.
\end{abstract}





\section{Introduction}
\IEEEPARstart {P}{arking} in urban areas is a challenging task \cite{khalid2021smart}. The clueless cruising for roadside parking can lead to congestion and wasted time and fuel. The urban parking pain arises from 1) increased car ownership, 2) limited availability of land for parking, and 3) lack of information on parking occupancy. A report by the International Energy Agency (IEA) \cite{a1} projects a 60\% boost in car ownership by 2070, and recent smart parking surveys \cite{b1} \cite{zulfiqar2023survey} reveal that urban parking facilities already took up to 31\% of urban lands. Given that the first two factors, i.e., high car ownership and limited land for parking, are beyond our control, it is crucial to develop a smart and effective solution to provide parking occupancy information. 

\par
In fact, most off-street parking lots, e.g., municipal car parks, already have systems to monitor parking occupancy either by entrance counters or fixed parking sensors \cite{shoup2013street} \cite{simicevic2023impact}. Though on-street parking has lots of advantages, the effect methods to detect the information of on-street are not well established \cite{biswas2017effects}. Thus, in this paper, we focus on discussing effective methods for providing on-street (roadside) parking information.

\par
In general, there are three types of techniques for getting parking information: fixed sensing, mobile sensing and data-based modeling.

\begin{itemize}
\item[$\bullet$] \emph{Fixed sensing:}
Fixed sensing is the most conventional technique among the three. It utilizes fixed-sensing devices, such as cameras, magnetic, infrared, ultrasonic, or radar sensors, to monitor changes in parking status \cite{b1}. 


\par 
This technique is quite reliable, but the deployment and maintenance costs associated with it can be prohibitively high. Firstly, covering an entire city with sensors would require an enormous number of sensing and related devices. The ratio of required sensors to the number of parking spaces ($\psi$) is equal to or greater than 1, i.e., $\psi \geq 1$. Secondly, the installation, replacement, and maintenance of these devices on municipal roads are  costly and complex.

\end{itemize}

\begin{itemize}
\item[$\bullet$] \emph{Mobile sensing:}

In contrast, mobile sensing uses sensors on vehicles to detect parking, offering easier operation and reduced maintenance compared to fixed sensing. It requires fewer sensors, as they scan spaces while vehicles move randomly through the city, significantly lowering the sensor-to-space ratio ($\psi \ll 1$), as shown in Figure 1.

\IEEEpubidadjcol
In \cite{b19}, the authors demonstrate that Crowd Sensing Intelligence (CSI) 
leverages diverse sensors for data collection but faces a trade-off in mobile sensing: higher detection accuracy demands more sensors on vehicles, increasing the required number of sensing units. In \cite{zhang2020mobile}, they design an intelligent mobile charging control mechanism for electric vehicles (EVs), which can be integrated with mobile sensing.

\end{itemize}

\begin{figure*}[htbp] 
\centering 
\includegraphics[width=0.85\textwidth]{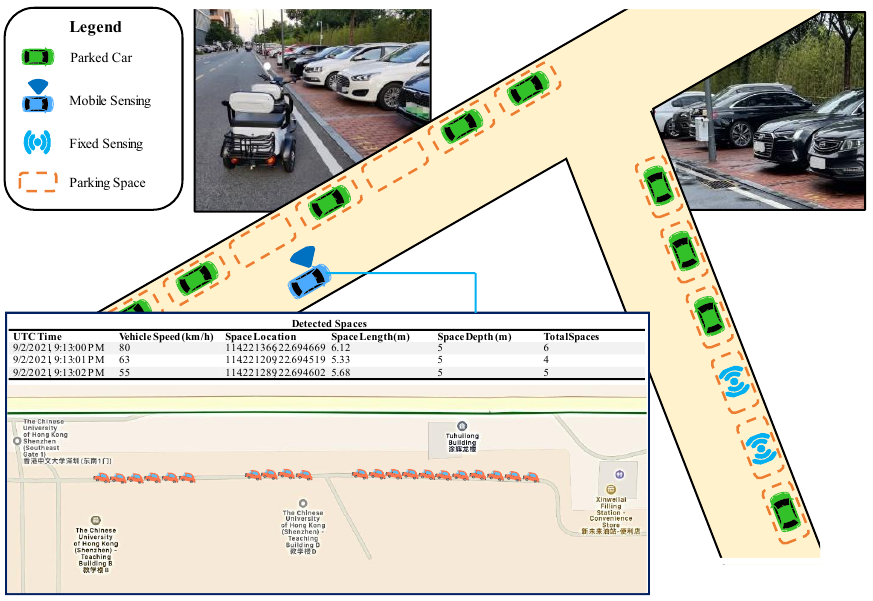}
\caption{Working scenarios of mobile sensing and fixed sensing techniques in monitoring parking status} 
\end{figure*}

\begin{itemize}
\item[$\bullet$] \emph{Data-based modeling:}
Data-based modeling involves the use of mathematical machinery and machine learning algorithms to predict available parking spaces \cite{b6}. The prediction model can be refined by using historical parking data, drivers' parking behavior, or parking demands. The performance of the prediction model can be further improved by considering additional factors such as time, weather, public holidays, and events. However, the inclusion of these factors increases the computing cost, and thus, complexity and performance are a trade-off in data-based modeling.
\end{itemize}

\par
Mobile sensing systems offer a cost-effective solution for on-street parking detection, further enhanced by crowd-sourcing techniques \cite{yi2015toward, liu2019unitask, zhang2014free}. In our previous work \cite{b7}, we deployed ultrasonic sensors on buses and taxis to monitor citywide roadside parking availability, using a supervised learning algorithm to determine parking status. We also introduced the Regular-Normal (R-N) algorithm \cite{b82} to improve crowdsensing accuracy. However, three key challenges remain unresolved.

\begin{enumerate}
\item 
 {
While mobile sensing provides a cost-effective solution for on-street parking detection, improving the detection accuracy between two consecutive detection intervals remains a challenge.
}
\item
 {
External factors like traffic and weather conditions can significantly influence parking patterns \cite{zhang2021periodic, yang2021truck}. A static parking detection approach fails to capture these urban dynamics, resulting in a gap between predictions and actual availability.}
\item
 {
Conventional parking algorithms are typically assessed using numerical metrics like the recall rate \cite{zhu2021driver}, overlooking the crucial aspect of drivers' parking decisions. Thus, a more comprehensive and practical assessment scheme is required.
}
\end{enumerate}

 {
We propose a Dynamic Gap Reduction Algorithm (DGRA), based on the integrated learning and optimization framework \cite{elmachtoub2022smart}, to address the challenges 1) and 2). DGRA is designed to improve the detection accuracy of the crowdsensing solutions. For example, the predictions between two consecutive detections of the R-N algorithm \cite{b82} were made at the middle time point of  two adjacent detections. DGRA automates the determination of detection frequency and the prediction time point by solving a stochastic optimization problem, which ensures that both decisions are optimized to maximize detection accuracy.
The key features of DGRA include: dynamic enhancement of detection accuracy with an online algorithm, integration of external influencing factors (e.g., traffic conditions), and adoption of Integrated Learning and Optimization (reducing the significance of distribution assumptions).
}

For the challenge 3), we introduce a driver-side traffic-based model (DSTBM) to incorporate drivers' perspectives into the algorithm, thereby enhancing its robustness and relevance for real-world scenarios. 

We validate the effectiveness and adaptability of DGRA through three categories of experiments: 1) tests using data from open data platforms, 2) comparing the performance of SFpark with that of DGRA, and 3) real-world drive tests at The Chinese University of Hong Kong, Shenzhen (CUHK-Shenzhen). The results confirm our system's  cost-efficiency, attributed to its use of crowdsensing, while maintaining robust performance both in numerical metrics and practical urban settings.

\par 
The rest of this paper is structured as follows: Section II reviews techniques for predicting parking availability. Section III introduces DGRA and discusses real-world deployment challenges. Section IV and V detail the driver perspective model and drive tests at CUHK-Shenzhen. Section VI analyzes numerical results, comparing scenarios with and without the DGRA. Finally, Section VII concludes the paper, outlines current limitations and highlights potential research directions.


\begin{table*}[htbp]
\centering
\caption{Features of three categories of parking solutions}
\begin{tabular}{|c|c|p{9cm}|}
\hline
\textbf{Type} & \textbf{Authors/Project Name} & \textbf{Features} \\ \hline

\multicolumn{1}{|c|}{\multirow{5}{*}{\textbf{Fixed sensing}}} 
 & Jung et al. \cite{b9} & Laser radar-based street corner detection \\ \cline{2-3} 
 & \multirow{2}{*}{SFPark \cite{bsf}} & The wireless sensor network structure is adopted. The pilot deployment cost is high \\ \cline{2-3} 
 & \multirow{2}{*}{Jin et al. \cite{jin2016occupancy}} & Sensing by Proxy (SbP) as a new paradigm for occupancy detection \\ \cline{2-3}
 \hline
 
\multicolumn{1}{|c|}{\multirow{8}{*}{\textbf{Mobile sensing}}} 
& \multirow{2}{*}{Roman et al. \cite{b7}} & The mobile sensing approach can perform as well as the fixed system, but the number of sensing units is significantly smaller \\ \cline{2-3} 
& \multirow{2}{*}{Bock et al. \cite{br13, br14}} & Consider multiple factors, such as road segments, taxi transit frequencies, and fleet sizes \\ \cline{2-3}
& \multirow{2}{*}{Kong et al. \cite{kong2018iot}} & An integrated auction and market design method for parking space sharing and allocation \\ \cline{2-3}
& \multirow{2}{*}{ParkNet \cite{br11}} & Reduce GPS error with the environmental fingerprinting approach \\
\hline
 
\multicolumn{1}{|c|}{\multirow{9}{*}{\textbf{Data-based modelling}}}  
 & \multirow{2}{*}{Rajabioun et al. \cite{b6}} & Parking guidance and information (PGI) systems using multivariate spatiotemporal models \\ \cline{2-3} 
 & \multirow{2}{*}{Kopecky et al. \cite{b13}} & Combine linked data and extra information, such as events and services, to make predictions more sensible \\ \cline{2-3} 
 & \multirow{2}{*}{Kim et al. \cite{kim2019parking}} & Minimize parking expenses and balance demand among public and private parking lots (PLs) \\ \cline{2-3}
 & \multirow{3}{*}{An et al. \cite{an2020parking}} & Destination privacy-preserving online parking sharing (DPOPS) incentive scheme that addresses urban congestion and illegitimate parking \\ \cline{2-3}
 \hline
\end{tabular}
\end{table*}


\section{Related Work}
This section reviews the solutions that provide on-street parking information. The main features of those solutions are summarized in Table I. 


\subsection{Fixed sensing}

\par 
Jung \emph{et al.} \cite{b9} proposed a method for scanning parking spaces by using laser-based radars. They developed a novel corner-detection strategy, including the detection of both rectangular and round corners, to enhance the detection accuracy. The high accuracy of a laser radar-based parking solution is remarkable, but the high cost of laser sensors limits the scalability of such systems.

\par

\par 

\par 
In the U.S.\ parking project SFpark \cite{bsf} by the San Francisco Municipal Transportation Agency, 8000 parking spaces were equipped with 11711 magnetometer sensors for periodically collecting and broadcasting information about the availability of parking spaces so that drivers can save time in cruising for parking spaces and congestion can be reduced.

\par
 Jin \cite{jin2016occupancy} introduces Sensing by Proxy (SbP) as a sensing paradigm for occupancy detection, leveraging proxy measurements like temperature and CO2 concentrations. The proposed framework employs constitutive models to capture the effects of occupants on indoor environments, enabling sensor fusion of multiple environmental parameters. Experimental results demonstrate the effectiveness of SbP in accurately inferring the number of occupants and its potential to significantly reduce ventilation energy consumption while maintaining occupant comfort.

\par 


Fixed sensing-based parking solutions, recognized for their precision and stability, are prevalent due to their capacity to promptly update parking status. Despite this, the substantial installation expenses and maintenance requirements hinder their widespread urban deployment. As fixed sensing is beyond this paper's scope, we direct readers to a comprehensive survey \cite{b1} for further insight.


\subsection{Mobile sensing}
Mobile sensing-based parking solutions eliminate the need to equip individual parking spots with sensors. The ParkNet model \cite{br11} endeavors to construct a real-time parking map by utilizing ultrasonic sensors and GPS units to acquire information regarding already parked vehicles. Additionally, environmental fingerprinting is used to enhance the precision of GPS. The model demonstrates an accuracy of over 90\% in the developed occupancy map, using the 500-mile parking data collected from test runs over two months.

\par 
Roman \emph{et al.} \cite{b7} developed a mobile sensing scheme as an alternative to the traditional fixed sensing approach, where the sensors are installed on the passenger side of a vehicle for measuring the distance from the vehicle to the nearest roadside obstacles. They evaluated their crowdsensing solution against that of a fixed sensing system obtained in Surrey in the UK. The obtained results showed that the mobile sensing approach could perform at an accuracy level similar to that of the fixed sensing approach even when the number of sensing units is significantly smaller.

\par 

\par 
Bock \emph{et al.} \cite{br13} simulated the sensing coverage of roadside parking by down-sampling the parking data from the SFpark project \cite{br11}. Assuming that a fleet of taxis equipped with sensors were capable to detect the available roadside parking spots, Bock \emph{et al.} \cite{br14} estimated the sensing coverage of different probing taxis based on their moving trajectories. They further investigated the suitability of the taxis to crowdsense on-street parking availability by considering multiple factors, such as road segments, taxi transit frequencies, and fleet size. The obtained results showed the crowdsensing of parking occupancy via taxis as a promising alternative to the expensive fixed-sensing based solutions. Kong \cite{kong2018iot} introduces an integrated auction and market design method for parking space sharing and allocation, utilizing a cloud platform enabled by the Internet of Things. The system employs price-compatible top trading cycles and chains  mechanism for private parking spaces and a one-sided auction for public spaces. Experimental results demonstrate that these mechanisms are effective.

\par 


\par 
In the mobile sensing paradigm, a single moving sensor can monitor multiple parking spots, not just individual ones, significantly minimizing sensor count through reuse. This crowdsensing technique gathers parking data directly from drivers, thereby offering dependable parking information through crowd-derived data analysis.


\subsection{Analytical modeling}
Data-based modeling involves making predictions using external inputs, such as traffic conditions, historical parking data, or drivers' parking demands. Rajabioun \emph{et al.} \cite{b6} proposed a parking guidance and information (PGI) system with multivariate spatiotemporal models. The authors evaluated the model using data provided by SFpark \cite{bsf}, and demonstrated the its effectiveness under different scenarios.



\par 
Kopecky \emph{et al.} developed an application scheme based on linked data for helping drivers in finding parking spots \cite{b13}. The application also integrates some additional data sources, such as municipal events and services, to make the predictions more sensible.

\par 
Kim et al. \cite{kim2019parking} addresses the issue by formulating a parking assignment problem aiming to minimize parking expenses and balance demand among public and private parking lots (PLs). Using a mixed-integer linear programming, the proposed method improves parking utilization by 27.5\%. An et al. \cite{an2020parking} proposes a destination privacy-preserving sharing scheme to address urban congestion and illegitimate parking by sharing private parking spaces. The scheme formulates the problem as a social welfare maximization issue in a two-sided market, using threshold rules and nonlinear programming to match winners while protecting customer location privacy with the Laplace mechanism.

\par 

\par

\par 
The data-based modeling uses historical parking data to predict parking occupancy, which  is a balance between the cost and accuracy. Although it is statistically feasible, the stable prediction of specific parking spaces remains as a challenging issue.


\section{Dynamic Gap Reduction Algorithm for Crowdsensing Roadside Parking Spaces}
In this section, we introduce the Dynamic Gap Reduction algorithm (DGRA) for increasing the detection accuracy of crowdsensing parking solutions. DGRA is extended based on our previous work \cite{b82}.  {Subsection A illustrates the conceptual foundations of the Dynamic Gap Reduction Algorithm. Subsection B provides a detailed exploration of the algorithmic framework, which consists of two integral components: a predictive model and a stochastic optimization problem. Subsection C articulates the rationale for selecting the predictive model. Subsection D introduces the stochastic optimization problem. Lastly, Subsection E  discusses the real-world deployment of DGRA.}


\subsection{Example of detection error}
In this paper, we use '0' to represent an occupied parking spot and '1' to represent an empty parking spot, as shown in Fig. 2. This notation follows the binary coding principle, where '0' denotes 'no', i.e., no free parking space, and '1' signifies 'yes', i.e., a free one. 
The parking and detection process is depicted in a timeline format, as shown in Fig. 2(a). 

In this crowdsensing method, the status of parking spaces is updated each time a detection car passes by. The timeline uses two colors to show different statuses: blue for a free parking space ('1') and green for an occupied one ('0'). When two rectangles of different colors intersect, it indicates a change in parking status. For instance, as seen in Fig. 2(a), a change from blue to green indicates that a vehicle has taken the parking space.

\begin{figure*}[!t] 
\centering 
\includegraphics[width=0.9\textwidth]{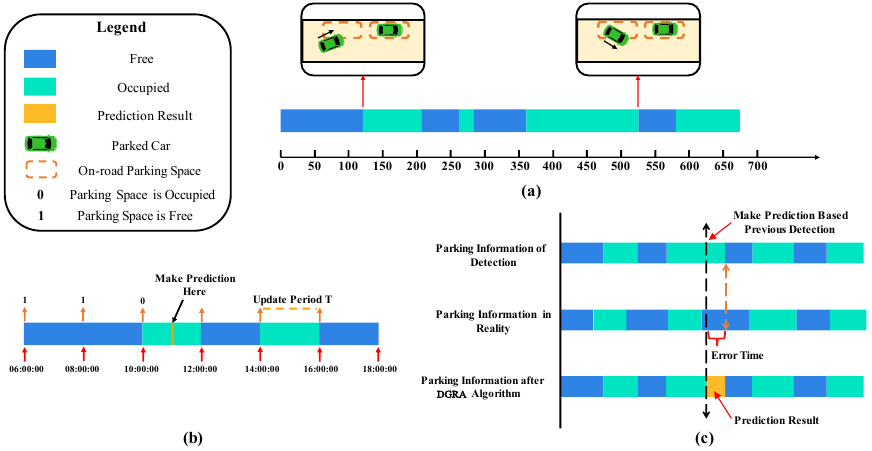}
\caption{Detection Error between two detections. (a) The process of entering or exiting a parking space, (b) The process of extra prediction between two detections, and (c) The benefit of DGRA prediction between two detections} 
\end{figure*}

\par 
In Fig. 2(b), the red arrow indicates the moment when the parking status is updated after the detection vehicles pass by the parking regions. The update frequency can be determined by measuring the time difference between two adjacent red arrows. For instance, the update frequency is once every two hours. However, an inaccurate prediction may occur because the parking status between two red arrows (a rectangle) is only checked at the first arrow point. Therefore, the detection result may be misleading if the parking status changes at any time instance between two adjacent arrows. Intuitively, increasing the update frequency can improve the detection accuracy.

Guided by this observation, we can generate additional predictions regarding the status of parking spots by considering historical detection results and factors influencing parking status, for example, an extra parking prediction between two detections, as shown in Fig. 2(b). 

As shown in Fig. 2(c), there are three detection timelines: parking information obtained through crowdsensing detection, parking information in reality, and parking information generated by applying DGRA. By comparing the first two timelines, it becomes apparent that inaccurate detection  can occur. To mitigate these errors in crowdsensing, the extra prediction is made using DGRA at the time indicated by the dashed arrow. Thus, the detection accuracy can be increased.

 {
\subsection{The framework of the Dynamic Gap Reduction Algorithm}
As mentioned in the preceding subsection, our task involves making predictions based on historical detection results and current environmental factors such as weather conditions and traffic flow, along with other influences on parking behavior. In this section, we discuss the framework of the Dynamic Gap Reduction Algorithm.} 

\begin{figure*}[!t] 
\centering 
\includegraphics[width=0.9\textwidth]{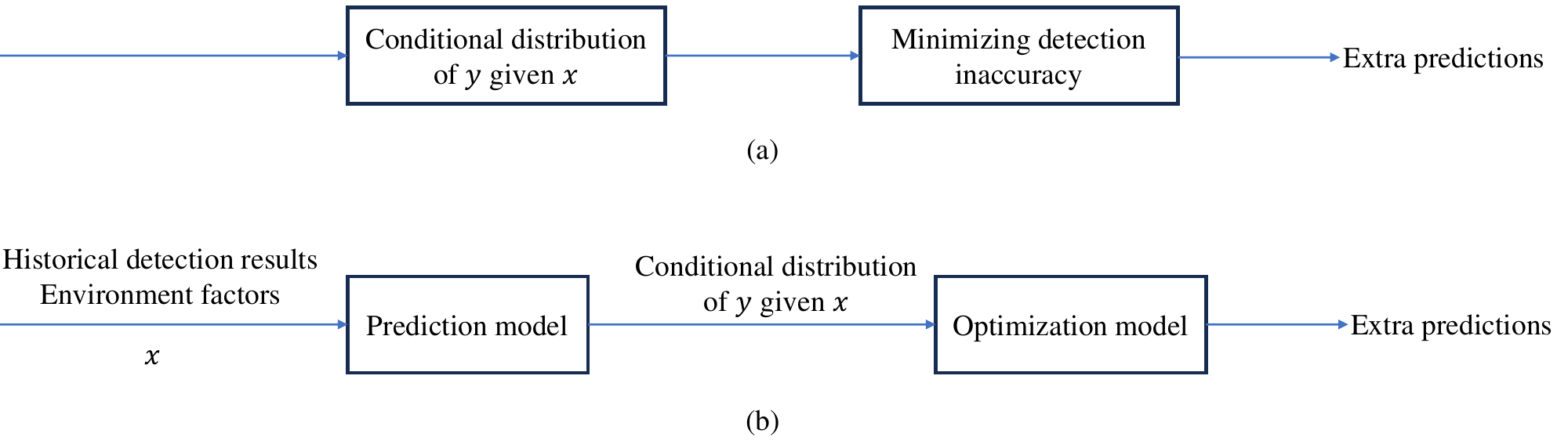}
\caption{The Framework of DGRA. (a) The process of DGRA in theory (b) The process of  DGRA in practice} 
\end{figure*}

 {
Let $x$ denote a multi-dimensional vector representing features/observations, encompassing variables such as weather conditions, traffic flow, and other factors influencing parking behavior.
Consider $y$ as a two-dimensional random variable. Specifically, $y_{1}$ signifies the duration a car remains parked in a parking spot, while $y_{2}$ denotes the duration a parking spot remains unoccupied.}

 {
Intuitively, as illustrated in the Fig. 3(a), optimization is performed considering the distribution of $y$ to minimize inaccuracies arising from the crowdsensing method. In spite that in practical scenarios, the distribution of $y$ is unknown, we can infer this distribution based on the features $x$. This can be regarded as the domain of contextual optimization \cite{sadana2024survey}. \cite{mahfooz2021context} \cite{rico2013parking} \cite{ul2019context} also mentioned the concept of contextual information in parking problem. However, they do not consider adapting it into the mobile sensing setting.}

 {
The DGRA consists of two primary components, as depicted in the Fig. 3(b). The first part is a predictive model that takes the feature vector $x$ as input and produces the conditional distribution of $y$ given $x$, where $y$ is a two-dimensional vector. The second part is a stochastic optimization model, determining the predicted time point and outcomes (i.e., whether a parking space is available at the predicted time point). In this context, we employ the Integrated Learning and Optimization (ILO) method  from \cite{elmachtoub2022smart}.}

 {\subsection{Kernel choice of random variable $y$}
In this section, we will   present the selection of two random variables, namely $y_{1}$ and $y_{2}$. The probability density function of random variable $y_{i}$ is represented by $f_{i}(\cdot)$, and $F_{i}(\cdot)$ stands for the cumulative distribution function of $y_{i}$.}

\begin{figure*}[htbp] 
\centering 
\includegraphics[width=0.67\textwidth]{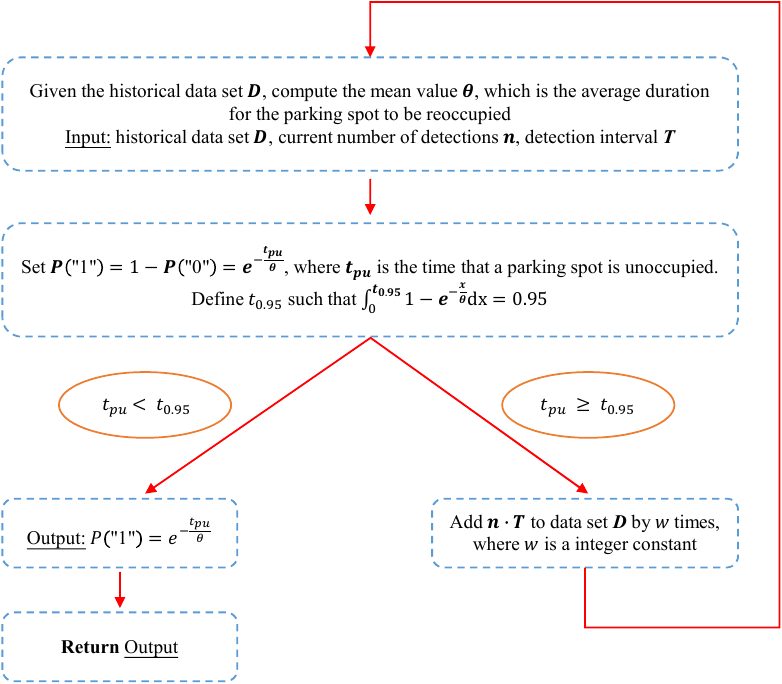} 
\caption{Flowchart of generating probability function P("1") for case 2} 
\end{figure*}

\begin{itemize}
\item[$\bullet$] \emph{Assumption of $y_{1}$} 

In practice, parking demands tend to concentrate in some specific time periods of the day. Additionally, the likelihood of a car being parked for either a very short or very long duration during those time periods is considerably low. Accordingly, we assume that the continuous random variable $y_{1}$ follows a normal distribution with mean value $\mu$ and variance $\sigma^2$, i.e., $y_{1} \sim N(\mu,\sigma^2)$.

\par 
When the value of $\delta$ is sufficiently small, $\int^{y_{1}+\delta}_{y_{1}-\delta}F(x)dx$ can be interpreted as the probability that a driver  keeps the car parked in the parking spot for a duration close to $y_{1}$. 

\item[$\bullet$] \emph{Assumption of $y_{2}$}

Let $\theta$ represent the mean time required for a parking spot to be re-occupied after the previous car leaves.  Then, we can use $F_{2}(y_{2}) = \int^{y_{2}}_{0}f_{2}(x)dx$ to represent the probability of a parking spot being occupied in the next moment after remaining empty for a duration of $y_{2}$.

\par 
1) When $y_{2}$ is not too large (specifically, not significantly larger than $\theta$), the probability function $F_{2}(y_{2})$ is likely to behave as an exponential distribution with $\theta$ as a parameter, i.e., $F_{2}(y_{2}) = 1-e^{-\frac{y_{2}}{\theta}}$.

\par 
2) If $y_{2}$ is much larger than $\theta$, it is an indication that parking demands are much later than usual. In such cases, we assume that some external factors, such as inclement weather or road work,  affect parking activities. Consequently, the probability function should be adjusted to reflect this potential abnormality, resulting in a longer time required for the probability to increase. 
Specifically, suppose $y_{2,i} = i\cdot T + t$, where $ 0\leq t < T$, and the modified probability function $F_{2}(y_{2})'$ must satisfy the following three conditions:
 \begin{align*}
    \textbf{(i)}\quad & F_{2}(y_{2, i})' < F_{2}(y_{2, i})\text{,\enspace for any~} y_{2} > 0, \enspace i\geq0 \\
    \textbf{(ii)}\quad & F_{2}(y_{2, 1})'> F_{2}(y_{2, 2})'> ... > F_{2}(y_{2, n})'\\
    \textbf{(iii)}\quad & F_{2}(y_{2, 1})'-F_{2}(y_{2, 2})' \leq F_{2}(y_{2, 2})'-F_{2}(y_{2, 3})' \\
    & \leq ... \leq F_{2}(y_{2, n-1})'-F_{2}(y_{2, n})'.
\end{align*}


\end{itemize}

 {Condition (i): External factors such as inclement weather or road construction can reduce the likelihood of parking activity. Therefore, the actual probability of a parking spot being occupied in the next moment should be lower, implying that for the same $y_{2}>0$, $F_{2}(y_{2},i)^'<F_{2}(y_{2},i)$.}

 {Condition (ii): The duration for which a parking spot remains unoccupied positively correlates with the abnormal impact's severity. When the abnormal factor has a high impact on parking activity, the unoccupied duration is longer. Thus, we consider the unoccupied duration as posterior information that reflects the severity of the abnormal impact. As the severity increases, the probability of a parking spot being occupied in the next moment decreases. In other words, an increase in unoccupied time results in a decrease in $F_{2}(y_{2},i)^'$, such that $F_{2}(y_{2},1)^'>F_{2}(y_{2},2)^'> \dots >F_{2}(y_{2},n)^'$.}

 {Condition (iii): As the duration of unoccupied time increases, the rate of decrease in $F_{2}(y_{2},i)^'$ should generally be faster, indicating a steeper slope of the decrease. Therefore, $F_{2}(y_{2},1)^'-F_{2}(y_{2},2)^'\leq F_{2}(y_{2},2) )^'-F_{2}(y_{2},3)^'\leq \dots \leq F_{2}(y_{2},n-1)^'-F_{2}(y_{2},n)^'$.}



\begin{table*}[!b]
\centering
\caption{ {Implementation Cost Estimation}}
\label{tab:cost_estimation}
\begin{tabular}{|c|c|c|}
\hline
\textbf{Key Components} & \textbf{Function} & \textbf{Unit Price} \\ \hline
Ultrasonic sensor & Measure space depth &  ¥4.4 (\$0.63) \\
Raspberry Pi Zero & Process data & ¥105 (\$15) \\
GPS module & Provide geo-location data & ¥72 (\$10.3) \\
4G module & Data transmission & ¥69 (\$9.9) \\ 
Cloud infrastructure & Computation and storage  & ¥15/month (\$2.1/month) \\ \hline
\multicolumn{2}{|r}{\textbf{Subtotal for one unit:}} & \textbf{¥265.4 (\$37.9)} \\ \hline
\end{tabular}
\end{table*}

 {
\subsection{The stochastic optimization problem}
Let $z$ be a two-dimensional decision variable in the optimization problem. $z_{1}$ indicates the predicted time point of DGRA, constrained within the interval $[0, \infty]$. Meanwhile, $z_{2}$ represents the predicted parking results, where '0' indicates an occupied parking spot, and '1' denotes an empty one.
Let $P(z_{2}|z_{1})$ be the probability that a parking spot is in state $z_{2}$ given the predicted time $z_{1}$. For example, $P(0|5)$ denotes the probability of the parking spot being occupied at time 5. Similarly, $P_{y|x}(z_{2}|z_{1})$ represents $P(z_{2}|z_{1})$ within the context of the conditional distribution $P_{y|x}$. Then, the stochastic optimization problem is formulated as follows:
\begin{align*}
\max_{z} & \, P_{y|x}(z_{2}|z_{1})\\
\text{s.t.} & \, z_{1} \geq 0 \\
& \, z_{2} \in \{0,1\}
\end{align*}}

 {
The computation of $P_{y|x}(z_{2}|z_{1})$ follows specific procedures that vary depending on two different cases.}

 {
\textit{Case I:} If the latest detection of parking occurred at $t_{1}$ with a result of '0', and for a given $z_{2}$, DGRA aims to predict the parking status at time $t_{1}+z_{2}$. Let $m$ represent the number of '0's from $t_{0}$ to $t_{1}$, obtained from historical detection results. Let  $T_{i}$ be the time duration between the i-th and (i+1)-th detection results of '0' among the $m$ detections. Then,
\begin{dmath}
P_{y|x}(0|z_{2})=\int^{\frac{2m-1}{2}z_{2}+\delta}_{\frac{2m-1}{2}z_{2}-\delta}f_{1}(x)dx+\\(1-\int^{z_{2}+\delta}_{z_{2}-\delta}f_{1}(x)dx)(\sum^{m-2}_{k=0}\int^{\frac{2k-1}{2}T_{k}+\delta}_{\frac{2k-1}{2}T_{k}-\delta}f_{1}(x)dx)~,
\end{dmath}
\begin{dmath}
    P_{y|x}(1|z_{2})= 1 - P_{y|x}(0|z_{2})
\end{dmath}}

\par 
 {The computation of $P(0)$ comprises two components: the first part indicates the probability of the parking space being empty, while the second part denotes the probability of the parking space being occupied but becoming empty within the time interval of $2\delta$.}

\par 
 {\textit{Case II:} If the parking information at time $t_{1}$ is '1', indicating that the parking spot is currently unoccupied, and considering that the last car vacated the spot at time $t_{0}$, let $n$ represent the count of '1's observed from $t_{0}$ to $t_{1}$. In this scenario, two sub-cases arise as a part of the kernel of $y_{2}$. To delineate these sub-cases, the time parameter $\alpha_{0.95}$ is defined to satisfy the condition $\int^{\alpha_{0.95}}_{0}h(x)dx = 0.95$, where $h(x) = 1-e^{-\frac{x}{\theta}}$.}

\par 
\textbf{Sub-case (1)} $y_{2}\leq \alpha_{0.95}$: 

\par 
The probability $P_{y|x}(0|z_{2}) = 1-e^{-\frac{y_{2}}{\theta}}$. So, 
\begin{equation}
P_{y|x}(1|z_{2}) = 1-P_{y|x}(0|z_{2}) = e^{-\frac{y_{2}}{\theta}}.\label{eq2} 
\end{equation}

\par 
\textbf{Sub-case (2)} $y_{2}>\alpha_{0.95}$: 

\par 
In  Sub-case (2), we need to modify the probability function as defined in Sub-case (1) in two steps, as shown in Fig. 4.
Recall that $\theta$ represents the mean time required for a parking spot
to be re-occupied after the last car leaves it. Now we consider $n$ as newly gathered data and add it to the dataset $w$ times, where $w>1$ is an integer constant. This step ensures that the change in the mean duration $\theta$ is significant, as adding $n$ data only once may result in a negligible change, given the large size of the dataset. Consequently, we obtain a new exponential distribution with the modified mean duration constant, $\theta'$. Finally, 
\begin{equation}
    P_{y|x}(1|z_{2}) = 1-P_{y|x}(0|z_{2}) = e^{-\frac{y_{2}}{\theta'}}.
\end{equation}  

\par 
It is easy to verify that the modified probability function $P_{y|x}(0|z_{2})$ satisfies the three properties specified  in Assumption 2. Hence, it is straightforward that $P_{y|x}(1|z_{2})$ also satisfies the same assumption.

\par 
It's worth noting that in both cases, if any change in the parking status is detected during the latest detection period (e.g., from '0' to '1' or vice versa), the algorithm will reset the starting time to the latest detection time. Additionally, as the parking status has changed, the computation procedure will also change.

 {
\subsection{Real-world deployment of DGRA}
}


 {
1) Data Privacy: In the real-world deployment, ensuring data privacy and security is of utmost importance. Inspired by the security framework for crowdsensing solutions \cite{n1}, the following strategies for enhancing data privacy can be considered.}

 {
\textbf{\emph{Anonymization:}} DGRA can benefit from integrating pseudonymous techniques. By assigning pseudo IDs to crowdsensing nodes, the system can maintain ID and location privacy but ensure secure   contribution tracking and reward distribution.}

 {
\textbf{\emph{Encryption:}} The adoption of location-based symmetric key generation and proxy encryption can  bolster data confidentiality without relying on a central authority.}


 {
2) Data Integration and Processing: In terms of data integration and processing, our algorithm examines the sonar trace's structure to distinguish between parked cars and road clutter. The analyzed data, which includes identified parked vehicles, available parking spaces, GPS coordinates, vehicle speed, and timestamps, is then transmitted to a cloud server for further processing. Key considerations during this procedure are discussed as follows.}


 {
\textbf{\emph{Selection bias:}} Crowdsensing is susceptible to selection bias and error. Offering incentives is a good way to foster participation from a more diverse cohort and more comprehensive data to mitigate this issue.}

 {
\textbf{\emph{Data synchronization:}}  Combining data from multiple vehicles can introduce data synchronization challenges. To address this, algorithms such as decision-level data fusion \cite{wei2020decision} and correctness verification \cite{liu2020privacy} can be employed to merge data from various sources with decent accuracy.}

 {
3) Implementation Cost: The DGRA crowdsensing system incorporates key components, a micro-controller, a GPS module, a communication module and a cloud server for data processing and storage. To estimate the implementation cost of the DGRA algorithm, we use an example setup with ultrasonic sensors and a Raspberry Pi Zero, as detailed in Table \ref{tab:cost_estimation}. 
}




\vspace{0.3cm}


\section{Driver-Side and Traffic-Based Evaluation Model}
Drivers' behavior in evaluating parking solutions is important. To account for this factor, we develop a Driver-Side and Traffic-Based Evaluation Model (DSTBM)\footnote{It is built upon our previous publication \cite{b81} at the 2022 18th International Conference on Intelligent Environments (IE), where it received the Best Presentation Award. DOI: 10.1109/IE54923.2022.9826771}. It consists of a driver's decision tree and a simulation process, which provide a standardized tool for evaluating the accuracy of parking detection by integrating drivers' perspectives on whether to park or not and the actual traffic conditions. 


\begin{figure*}[!htbp] 
\centering 
\includegraphics[width=0.8\textwidth]{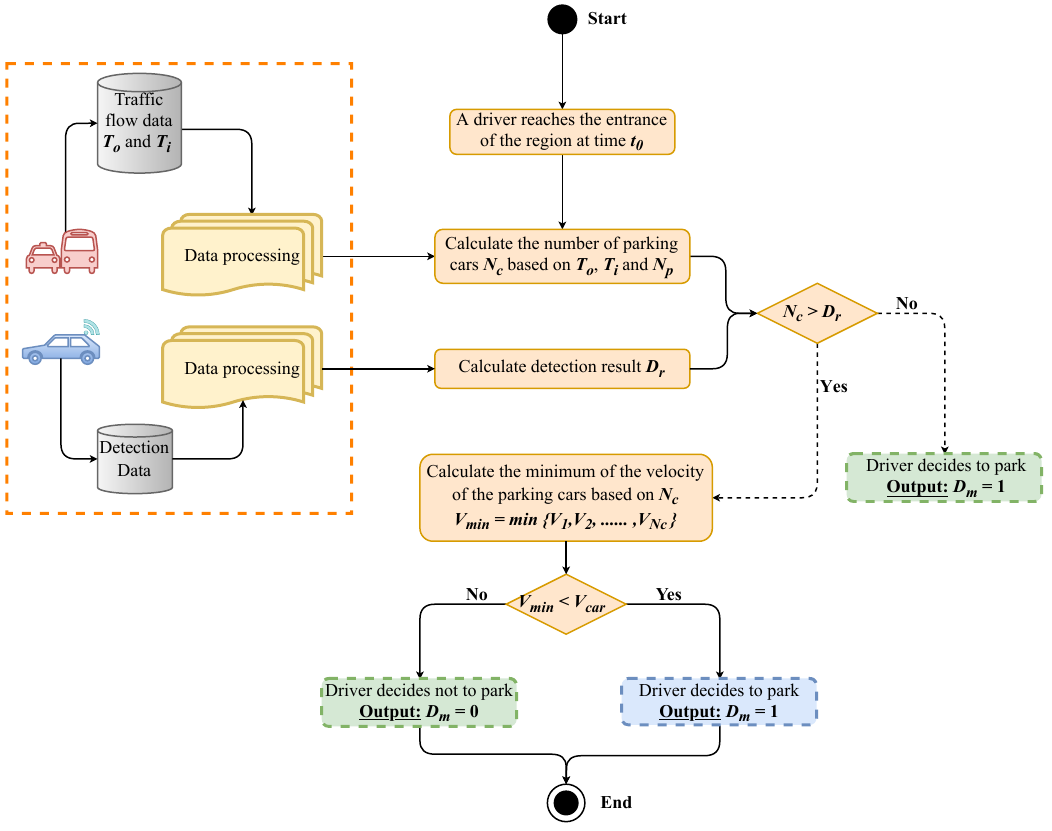} 
\caption{Drivers' decision model based on external factors}
\end{figure*}



\subsection{DSTBM Preliminary}
As discussed in Section II, various techniques exist for detecting the availability of on-street parking spaces, each with a specific approach for evaluating its detection accuracy. While some techniques focus on the temporal availability of parking spaces, others emphasize the spatial accuracy. Thus, a standard evaluation method is required to compare different parking detection solutions. To this end, we present DSTBM, which considers drivers' perspectives and actual traffic conditions in the parking region. DSTBM  is developed based on four assumptions.

\begin{itemize}
    \item Assumption 1: Each car passes through a parking region at a constant speed, and the difference between the speeds of any two cars is negligible. 
    \item  {Assumption 2: Drivers are patient and willing to wait for a parking spot to become vacant, rather than leaving the area to seek parking elsewhere.}
    \item Assumption 3: If a car is parked, its parking duration $y_{1}$ ($y_{1} \geqslant 0$) is considered as a random variable that follows a normal distribution having a mean of $\mu$ and variance of $\sigma^2$.  {This assumption, justified by the Central Limit Theorem \cite{clt} due to the model's large data size, leverages historical parking data from Shenzhen open data platform\footnote{\url{https://opendata.sz.gov.cn/data/dataSet/toDataDetails/29200_00403592}} to estimate $\mu$ and $\sigma^2$}. 
    \item Assumption 4: Drivers arrive at the parking region at a time interval of $T_{interval}$, which is treated as a continuous random variable that follows an exponential distribution with an arriving rate of $\lambda$. Since the drivers' behaviors are independent of $T_{interval}$, the memory-less feature of the exponential distribution is upheld.
\end{itemize}


\begin{figure*}[!t] 
\centering 
\includegraphics[width=0.88\textwidth]{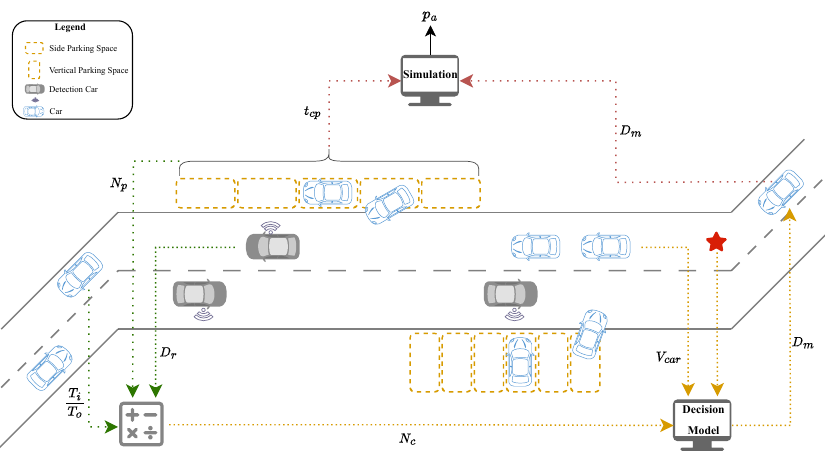}
\caption{Working flows of DSTBM} 
\end{figure*}


\subsection{Drivers' decision tree}
Upon arriving at the entrance of the parking region, drivers must take into account various factors before deciding whether to park their vehicles. The following section presents the drivers' decision model, which aims to predict their parking decisions.

\par 
The drivers' parking decision model is summarized in Fig. 5. The model takes a car's arrival time $t_0$ as input and generates the driver's parking decision $D_m$ as output. When a car arrives at the parking region, the model counts the total number of cars ($N_c$) cruising for parking spaces using traffic flow data. It then compares $N_c$ with the crowdsensing result ($D_r$, the number of available parking spaces detected by a mobile sensing system) to predict the drivers' parking decisions.  {If $N_c > D_r$, it indicates that there are insufficient parking spots in the parking region for all the cruising cars. However, individual drivers may still opt to park depending on the traffic conditions. When a driver's speed $V_{car}$ equals or exceeds the minimum speed of other cars ($V_{car} \geqslant V_{min}$), the driver will remain and compete with others for a parking spot ($D_m=1$). Otherwise, considering our assumption that $N_c > D_r$, any vehicles entering the parking region will inevitably compete for the limited parking spots. Consequently, vehicles with lower speeds have a low probability of securing a parking spot in this competitive environment. Therefore, we  incorporated a prioritization system in our model to favor faster-moving vehicles and increase their likelihood of finding an available parking spot.}

If $N_c\leqslant D_r$, it means that there are enough parking spaces to accommodate all the cruising vehicles. Thus, the driver will stay and get a parking spot ($D_m=1$).

DSTBM integrates the drivers' decision model and a  detection simulation process as shown in Fig. 6. In the simulation process, DSTBM takes the drivers' parking decision as input to obtain the prediction accuracy $P_a$. Note that the output $D_m=1$ of the drivers' decision model means that a driver decides to park. However, if the driver fails to park at the end, we consider the output as a wrong prediction (a false positive). Hence, the prediction accuracy $P_a$ of a parking solution can be calculated as true positive plus true negative as follows:

\begin{align*}
\text{P}_a=&\frac{\text{free space estimated \& driver actually parked}}{\text{total prediction}} + \\
&\frac{\text{no free space estimated \& driver actually not parked}}{\text{total prediction}}.
\end{align*}

\subsection{Illustration of the whole process}
The number of available parking spots ($N_p$), number of available parking spaces ($D_r$) detected by a mobile sensing technique, and traffic flows ($T_i/T_o$) on the parking street are updated continuously. The real-time number of cruising cars ($N_c$) in the parking region is calculated in terms of the above three variables and updated  in the drivers' decision model.

\par 
Suppose $G(T_{interval})$ is the probability density function of $T_{interval}$, which is the time interval of cars reaching the parking region. Hence, the probability that a car will enter the parking lot at the next moment is $\int^{t+\delta_1}_{t-\delta_1}G(x)dx$, where $\delta_1$ is a small number and $G(x) = \lambda e^{-\lambda x},\,x \geq 0$. Accordingly, a time sequence for the cars reaching the parking region can be generated.

\par 
Upon a car's arrival at the parking region (the red star in Fig. 6), the drivers' decision model is activated, which combines $N_c$ and $V_{car}$ to produce the output $D_m$ indicating whether the car will be parked or not. Then, the driver behaves according to the output of the decision model. Meanwhile, $D_m$ is transmitted to the simulation process as the input, which is combined with $y_{1}$ (the parking duration of a car) to determine the prediction accuracy.

\par 
According to \emph{Assumption 3}, the parking duration $y_{1}$ follows a normal distribution. So, the probability that the car is still parked in the parking space after $y_{1}$ is  $\int^{d+\delta_2}_{d-\delta_2}F(y)dy$, where $\delta_2$ is a small number and $F(y)=\frac{1}{\sigma\sqrt{2\pi}}e^{-\frac{(y-\mu)^2}{2\sigma^2}},~ y \geq 0$.

\par 
When the next car arrives at the entrance of the parking region, the same procedure is repeated. Therefore, using the real-time value of $y_{1}$, we can obtain the prediction accuracy of the parking solution. 

\subsection{Sensitivity Analysis of DSTBM}
\begin{figure}[htbp] 
\centering 
\includegraphics[width=0.5\textwidth]{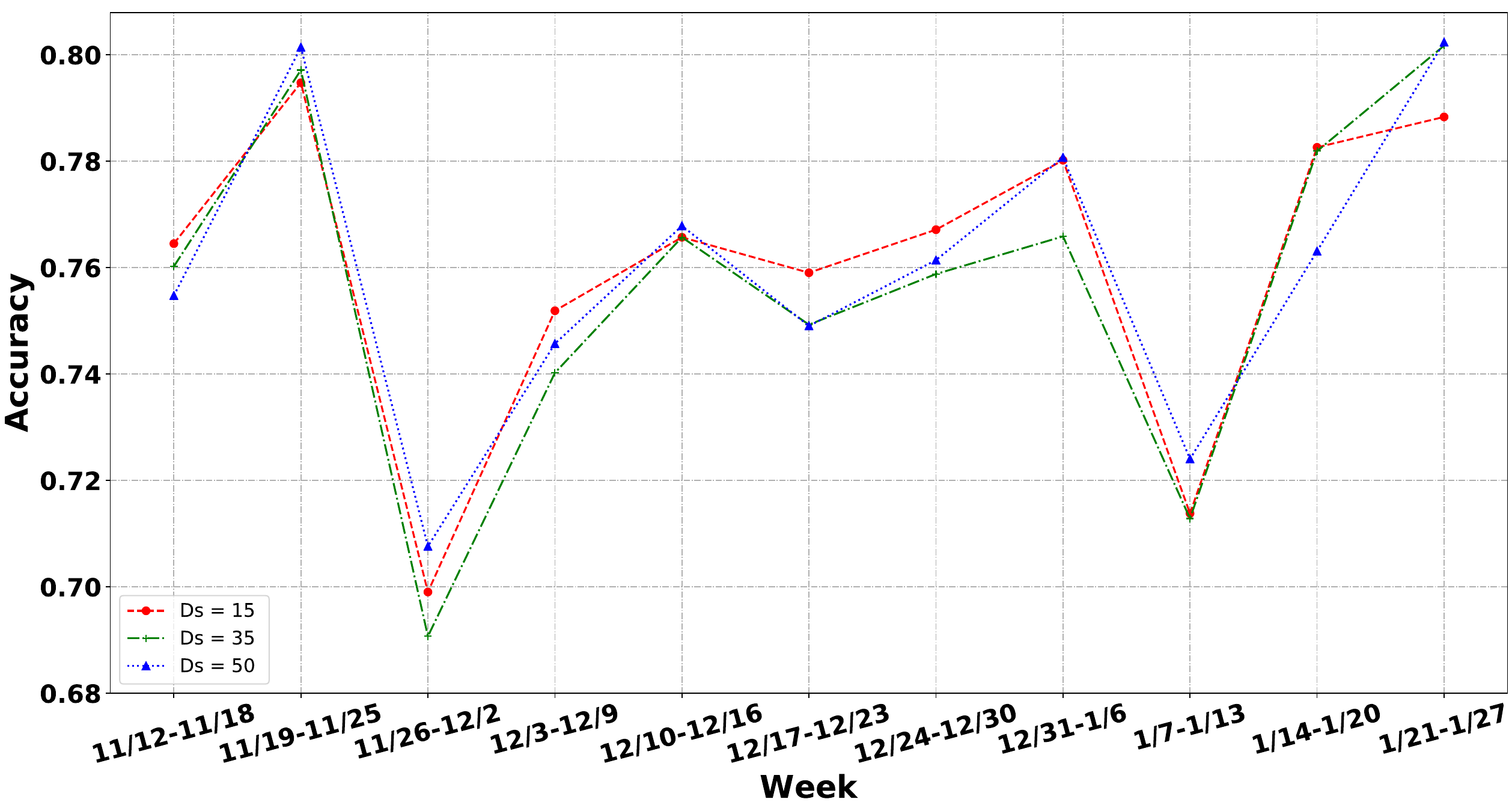} 
\caption{ {DSTBM with various $D_s$}} 
\label{sensitivity} 
\end{figure}

The sensitivity of DSTBM to the detection schedule parameter $D_s$ is evaluated and presented in Fig. \ref{sensitivity}. This parameter, crucial for determining the update frequency of parking occupancy status, directly affects the model's predictive accuracy. 

We conducted experiments on the dataset from the open data platform of Nanshan District of Shenzhen, China, spanning from September 1st, 2018, to January 1st, 2019, which  consists of 624,464 parking event records, with a network of 1735 sensors deployed for parking status detection \cite{b17}. Fig. \ref{sensitivity} demonstrates detection accuracy against time over several weeks, with lines representing different $D_s$ values: 15, 35, and 50 minutes. Generally, a 15-minute detection schedule (\( D_s=15 \), red line) achieves higher accuracy against other schedules, which is consistent with the expectation that more frequent updates improve prediction quality.

There are instances where \( D_s=50 \) outperforms \( D_s=15 \), likely due to the inherent variability in the real parking data. An extended version of DSTBM \cite{shi2024driver} addresses these fluctuations. However, the current version of DSTBM remains effective in capturing the necessary patterns for accurate predictions.


\section{Experimental Setup}
In this section, we present the practical implementation of the DGRA framework using ultrasonic sensors as a case study. A comprehensive description of the experimental setup is provided, including the evaluation metrics, equipment specifications, drive-test route, and the step-by-step experimental procedure.

\begin{figure}[htbp] 
\centering 
\includegraphics[width=0.43\textwidth]{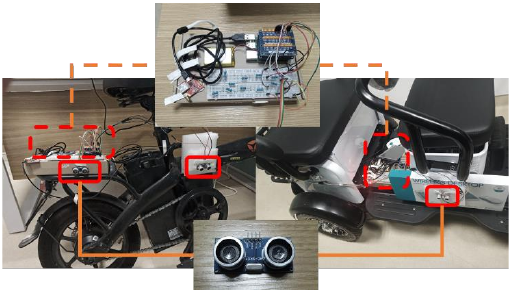} 
\caption{Mobile sensing vehicles} 
\label{sensing vehicles} 
\end{figure}

\begin{enumerate}
\item   { \emph{Evaluation indicator:}
The new indicator can be regarded as a composite of two factors: the accuracy of parking detection and information accuracy. The concept of detection accuracy is extensively elaborated throughout the paper. On the other hand, information accuracy refers to the percentage of instances in which drivers perceive the instructions provided by the system were consistent with the ground truth. This indicator holds greater significance compared to detection accuracy alone. It was inspired by a prominent topic in economics that combines prediction and optimization. Traditionally, the focus has been on minimizing prediction error as the primary performance criterion. However, this approach has limitations. Thus, the newly introduced decision error factor combines optimality and  prediction error into a unified measure, providing a more comprehensive performance evaluation.}

\item \emph{Experimental equipment:}
Two mobile sensing vehicles are deployed in this test, as shown in Fig. 8. Each vehicle is equipped with a  detection kit, containing an ultrasonic sensor and a micro-controller command system. In Fig. 8, the ultrasonic sensors are marked by solid-line rectangles and the command system by dash-line rectangles. 

\item \emph{Drive-test route:}
The experiment was conducted in the lower campus of CUHK-Shenzhen. As shown in Fig. 9, there are two parking regions along the drive-test route, Region A and Region B. All the parking units in those regions are perpendicular to the street. Region A has 20 parking units and Region B has 4 parking units, with each parking unit containing five individual parking spots. Hence, there are in total 120 roadside parking spaces. The detected vehicles started at the blue mark, drove along the red line and passed by Region A and Region B every 10 minutes. 

\begin{figure*}[!htbp] 
\centering 
\includegraphics[width=0.8\textwidth]{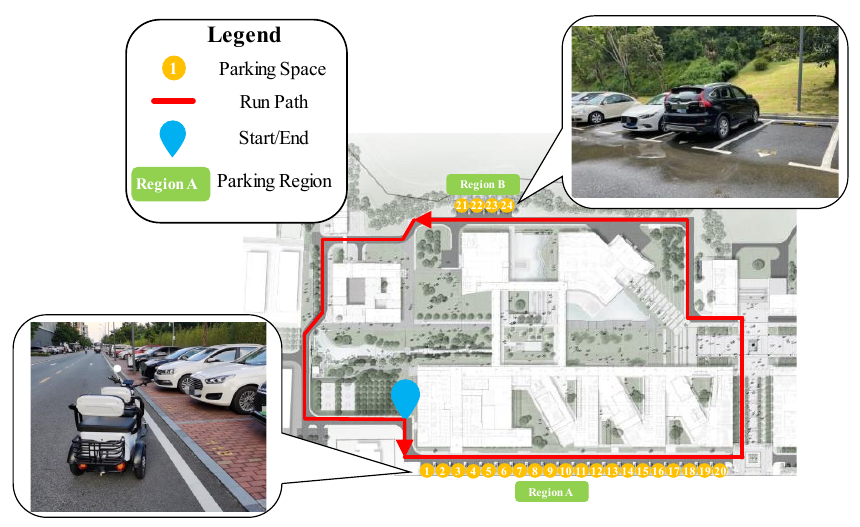} 
\caption{Configuration of the drive test in the campus} 
\end{figure*}

\item \emph{Experimental procedure:}
The experiment was conducted in the morning from 08:00 hours to 09:00 hours, and in the afternoon from 16:00 hours to 18:00 hours. The main reason for selecting the two time periods is that the parking occupancy changes frequently in these periods. The ultrasonic sensors scanned the parking occupancy when a detection vehicle drove by the parking regions. Then, the command system analyzed the collected data with/without applying  DGRA. Finally, the detection performance was evaluated by DSTBM.
\end{enumerate}



\section{Numerical Results}
 {
We employed three distinct scenarios to assess the performance of  DGRA. In Scenario I, we utilized historical parking data obtained from open data platforms in Shenzhen \cite{b17} and Shanghai \cite{b18} in China. This was done to illustrate the practical application of  DGRA.
In Scenario II, we present results based on data from SFpark, aiming to compare the algorithm's performance against other prominent projects.
In Scenario III, we conducted a real-world experiment using traffic and parking data collected from drive tests on the CUHK-Shenzhen campus.} 

\begin{figure*}[htbp] 
\centering 
\includegraphics[width=0.9\textwidth]{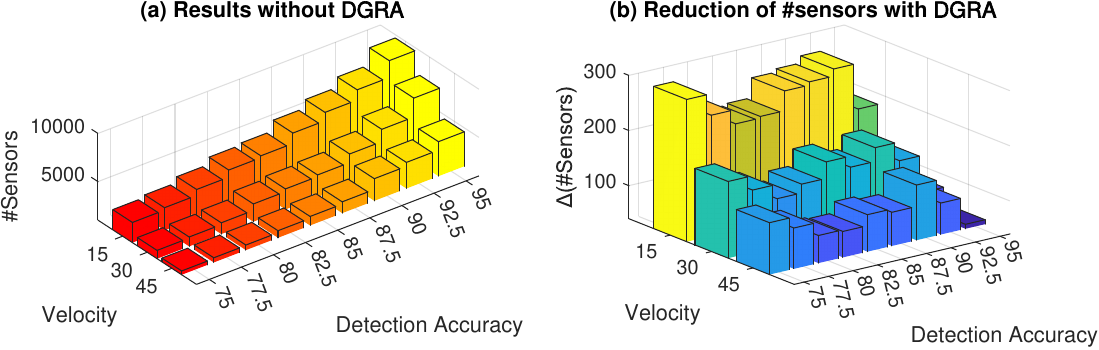} 
\caption{ {Visualization of the experiment results. (a) Without  DGRA, and (b) The reduction in number of sensors with  DGRA}}
\end{figure*}

\begin{table*}[!htbp]
\centering
\caption{Detection accuracy and number of sensors with DGRA}
\label{table2}
\renewcommand{\arraystretch}{1.05} 
\begin{tabular}{|c|c|c|c|c|}
\hline
\multirow{2}{*}{\textbf{Detection Interval (T)}} & \multirow{2}{*}{\textbf{Detection Accuracy}} & \multicolumn{3}{c|}{\textbf{Number of Sensors = 3286}} \\ \cline{3-5} & 
& \textbf{V = 15km/h} & \textbf{V = 30km/h} & \textbf{V = 45km/h} \\ \hline
15min     & 96.2\%     & 9646     & 7446    & 4568     \\ \hline
20min     & 93.9\%     & 6835     & 4384    & 3286     \\ \hline
25min     & 92.4\%     & 4689     & 2973    & 1743     \\ \hline
30min     & 88.7\%     & 3961     & 2172    & 1544     \\ \hline
\end{tabular}
\end{table*}

\begin{table*}[!htbp]
\centering
\caption{With and without applying DGRA}
\label{table3}
\renewcommand{\arraystretch}{1.1} 
\begin{tabular}{|ccccccc|}
\hline
\multicolumn{2}{|c}{\textbf{Detection Interval (T)}}                    & 15min  & 20min  & 25min  & 30min  & 35min  \\ \hline
\multirow{2}{*}{\textbf{Detection Accuracy}}   & Without DGRA  & 94.5\% & 90.2\% & 89.4\% & 79.5\% & 74.2\% \\ \cline{2-7} 
                                               & With DGRA     & 96.3\% & 93.8\% & 92.4\% & 88.7\% & 86.0\% \\ \hline
\multicolumn{2}{|c}{\textbf{Improvements After Applying DGRA}} & 1.8\%  & 3.6\%  & 7.5\%  & 9.2\%  & 11.8\% \\ \hline
\end{tabular}
\end{table*}

\subsection{Results of Scenario I}
In the method I, we assume that buses are equipped with ultrasonic sensors. The roadside parking spaces are divided into two groups: on-bus-route and off-bus-route parking spaces. The detection frequency is highly related to the number of sensing buses. 



\subsubsection{Results without DGRA}
The relationship between the detection accuracy and the number of sensors is shown in Fig. 10(a). It is obvious that the detection accuracy increases as the number of sensors or detection frequency increases.


\subsubsection{Results with DGRA}
Using the identical collection of open data ($g = 0.05$), while keeping the same detection accuracy, the number of required sensors can be reduced, as shown in Fig. 10(b). The corresponding data are presented in TABLE III. Setting the speed of the detection vehicle to $V = 45$ km/h, the number of sensors to 3286, and the detection frequency to $T = 20$ minutes, a detection accuracy of $93.9\%$ is obtained (TABLE III). The results indicate that the implementation of  DGRA  can improve detection accuracy. In comparison to the supervised learning algorithm presented in \cite{b7}, the proposed DGRA algorithm enhances detection accuracy by inserting additional predictions between two consecutive mobile scans.

\par 
To further investigate, we compared the results obtained with and without the DGRA algorithm ($g = 0.05$, $V = 45$ km/h) as shown in Fig. 11. The results demonstrate that  DGRA  significantly improves detection accuracy, particularly when the detection frequency is low (i.e. the number of sensors is small). Table IV presents a sampled comparison, indicating that  DGRA  performs better with larger sensing intervals.

\begin{figure}[htbp] 
\centering 
\includegraphics[width=0.48\textwidth]{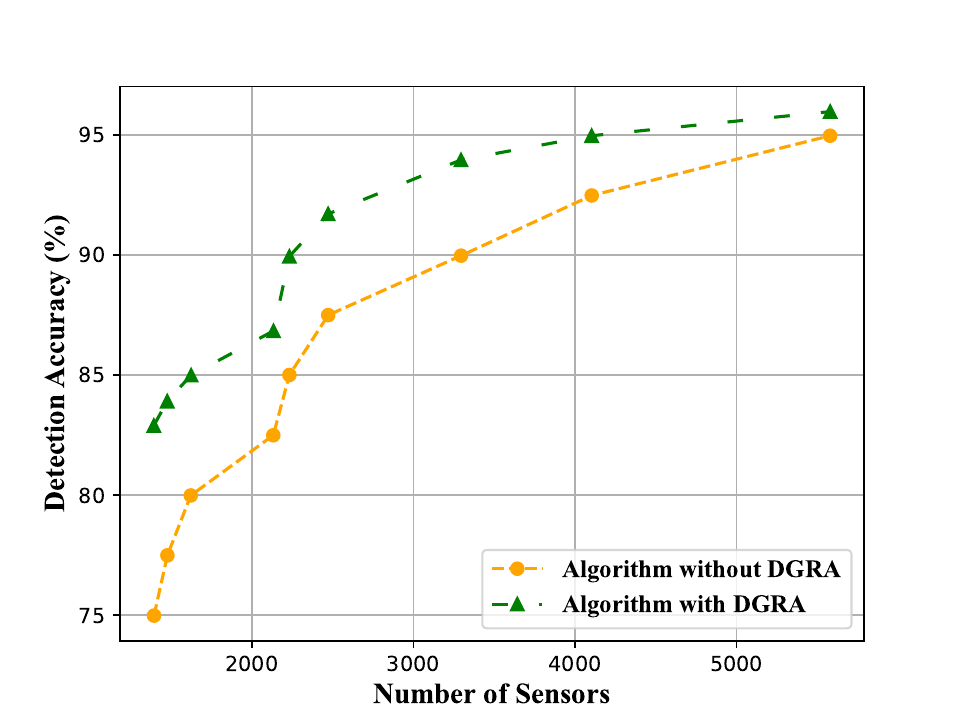} 
\caption{Accuracy against  {number of moving sensors} with $g=0.05$ and $V=45km/h$} 
\label{Fig.7} 
\end{figure}

 {
\subsection{Results of Scenario II}
In this scenario, we present results based on data provided by SFpark \cite{br13} to showcase the performance of DGRA. As illustrated in Fig. 12, several observations are obtained.
}

 {
\begin{itemize}
    \item As the detection frequency increases, the detection accuracy  improves in all three methods presented.
    \item In the high detection frequency range, SFpark outperforms the crowdsensing method. Conversely, in the low detection frequency range, SFpark performs less effectively than the crowdsensing method.
    \item DGRA improves the crowdsensing performance, surpassing that of SFpark  across the entire range.
\end{itemize}
}
\subsection{Results of Scenario III}
We aim to compare the detection performance before and after the implementation of DGRA. To achieve this, we conducted drive tests in the CUHK-Shenzhen campus. The results of the experiment are presented in Table \ref{tab:table34}.

\begin{table*}[!htbp]
\caption{Drive-test Results}
\centering 
\includegraphics[width=\textwidth]{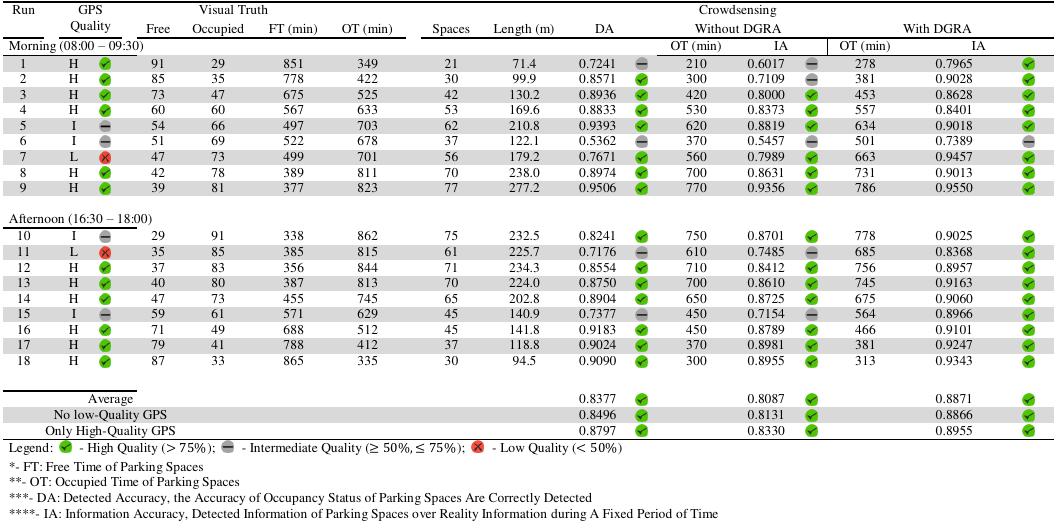} 
\label{tab:table34}
\end{table*}

\subsubsection{Notations in Table \ref{tab:table34}}
The algorithm was executed 9 times in the morning and 9 times in the afternoon. We also recorded GPS data as it influenced the spatial results of the parking spaces. The \emph{Visual Truth} denotes the ground truth of the parking status. The parking lot had a total of 120 spaces, and the number of free and occupied spaces are represented by \emph{Free} and \emph{Occupied} respectively. Additionally, we calculated the free time of the parking spaces, denoted by "FT," which sums up the time periods when the parking spaces remain unoccupied for more than 10 minutes. In contrast, \emph{OT} counts the time when the parking spaces remain occupied. In the crowdsensing column, \emph{Spaces} means the number of parking spaces detected by a mobile sensing vehicle. \emph{Length} indicates the total length of the detected parking spaces. Detected accuracy  (\emph{DA}) is the ratio of the number of occupied spaces detected to the number of spaces occupied in reality. Information accuracy  (\emph{IA}) is defined as the ratio of the detected time during which the parking spaces remained occupied to the actual time during which the spaces were occupied, over a fixed period of 10 minutes.


\subsubsection{Analysis of experimental results}
Table \ref{tab:table34} shows that the average detection accuracy (\emph{DA}) across all runs is approximately $84\%$, with a detection frequency of once every 10 minutes. As expected, poor GPS quality leads to a decrease in DA, indicating that GPS quality has a significant impact on detection accuracy.

\par 
We can also observe that information accuracy  (\emph{IA}) is enhanced upon applying  DGRA. The average \emph{IA} with  DGRA  is above 88\%, and it goes up to 90\% with high GPS quality. 

\begin{figure}[!t] 
\centering 
\includegraphics[width=0.5\textwidth]{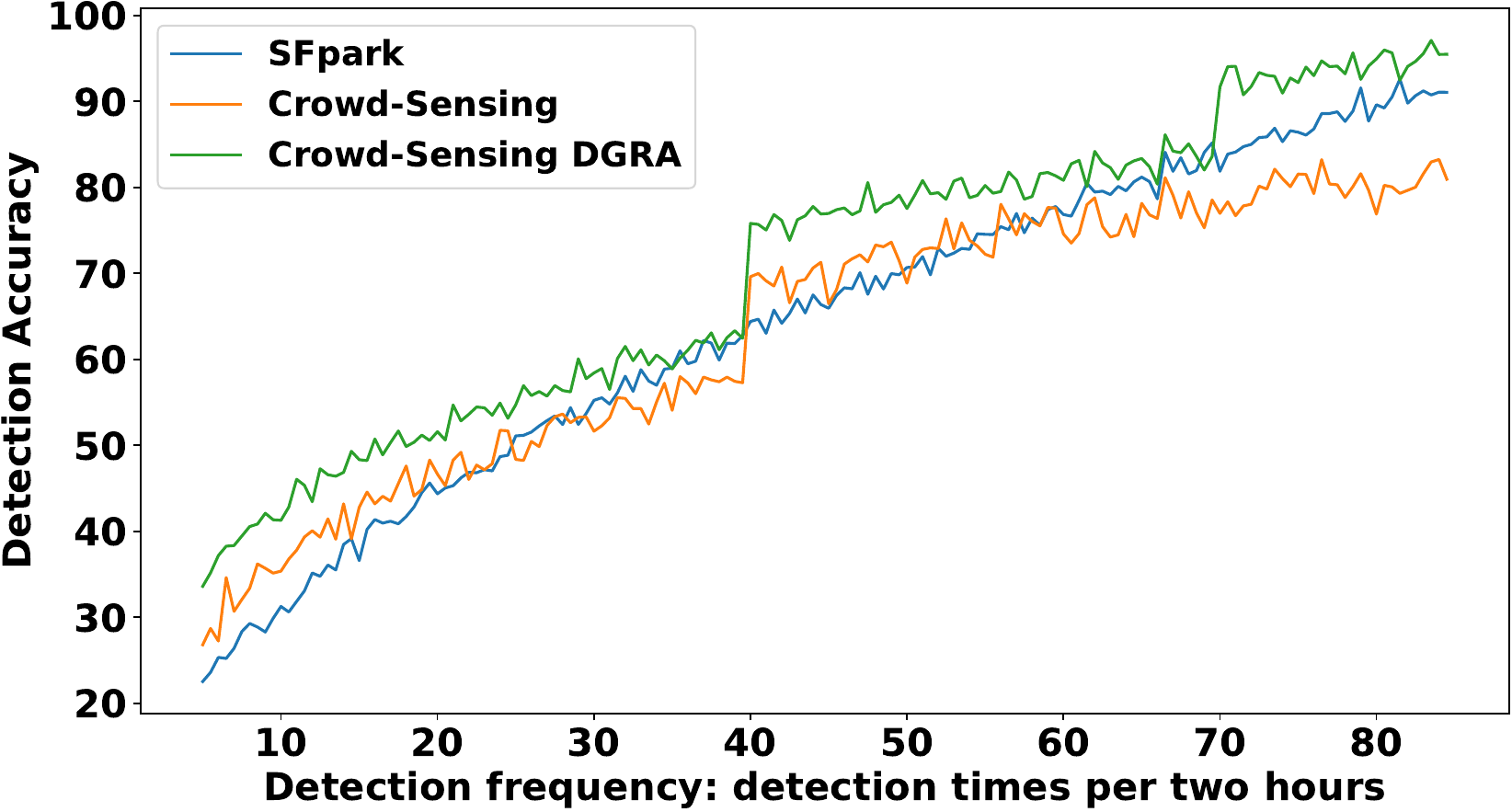} 
\caption{Comparing DGRA with SFpark} 
\label{Fig.10} 
\end{figure}

It is evident from our results that the implementation of DGRA leads to an improvement in IA, particularly when the accuracy is around $80\%$. This finding is consistent with our previous observation that the lower the initial accuracy rate, the better the performance we can achieve with  DGRA. Specifically, we found that  DGRA  resulted in an IA of above $82\%$ in nearly $88\%$ of runs.


\section{Conclusion and Future Work}
This paper presents Dynamic Gap Reduction Algorithm (DGRA) designed to improve the detection accuracy and algorithm generalizability of a crowdsensing parking solution. In addition, we introduce a novel evaluation method that takes into account drivers' perspectives, with the goal of achieving a more robust algorithm. Our study demonstrates that  DGRA enjoys a high level of cost-efficiency while maintaining robust performance both in numerical metrics and practical urban settings.

However, DGRA has certain limitations and potential failures that must be acknowledged. First, its effectiveness is highly dependent on the quality and granularity of historical parking data, with inaccuracies potentially leading to unreliable predictions. Second, DGRA's performance may be compromised in highly dynamic environments where parking patterns shift rapidly due to factors like municipal events, road closures, or traffic fluctuations. Lastly, the scalability of the algorithm could be impacted by device limitations and network connectivity issues when applied across diverse urban settings.

In future research, we plan to address these limitations and improve DGRA. Our focus will be on robustly handling historical parking data to mitigate inaccuracies. Additionally, we will strive to enhance the algorithm's adaptability and scalability across different urban environments.

\bibliographystyle{IEEEtran}
\bibliography{ref}

\begin{table*}[!htbp]
\caption{Parameters of DGRA}
\centering
\begin{tabular}{|c|l|}
\hline
\textbf{Notation} & \textbf{Meaning} \\
\hline
$0$          & representation of an occupied parking
spot \\
$1$        & representation an empty parking spot \\
$x$           & observations/features \\
$y_{1}$        & a random variable representing the duration a car remains parked in
a parking spot \\
$y_{2}$          & a random variable representing the duration a parking spot
remains unoccupied. \\
$f_{i} (\cdot)$       & the probability
density function of random variable $y_{i}$\\
$F_{i}(\cdot)$           & the cumulative distribution function of random variable $y_{i}$\\
$z_{1}$      & decision variable indicating the predicted 
time point of the dynamic gap reduction algorithm \\
$z_{2}$ & binary decision variable indicating the predicted parking results \\
$P_{y|x}$ & conditional distribution of $y$ given $x$\\
$P(z_{2}|z_{1})$ & the probability that a parking spot is in state $z_{2}$
given the predicted time $z_{1}$ \\
$P_{y|x}(z_{2}|z_{1})$ & $P(z_{2}|z_{1})$ within the context of the
conditional distribution $P_{y|x}$\\
\hline
\end{tabular}
\end{table*}

\begin{table*}[!htbp]
\caption{Parameters of DSTBM}
\centering
\begin{tabular}{|c|l|}
\hline
\textbf{Variable} & \textbf{Explanation} \\
\hline
\( c \)           & The number of all parking spaces in parking region \\
\( \rho \)        & Service intensity of the parking system \\
\( A \)           & The distribution of drivers arriving at the parking region \\
\( N \)           & The number of drivers arriving at the parking region (the same as \( N_a \) in the simulation process) \\
\( d \)           & Time matrix for drivers’ arrival and leaving \\
\( st/lt \)       & Arrival time / leaving time of drivers \\
\( T \)           & Total simulation time \\
$\mathbf{D_s}$      & Cruising schedule of detection vehicles \\
\hline
\end{tabular}
\end{table*}

\end{document}